# Analysis of the vulnerability of machine learning regression models to adversarial attacks using data from 5G wireless networks

Leonid Legashev
Research Institute of Digital Intelligent Technologies
Orenburg State University
Orenburg, Russia
silentgir@gmail.com

Artur Zhigalov
Research Institute of Digital Intelligent Technologies
Orenburg State University
Orenburg, Russia
leroy137.artur@gmail.com

Denis Parfenov
Research Institute of Digital Intelligent Technologies
Orenburg State University
Orenburg, Russia
parfenovdi@mail.ru

*Abstract*— This article describes the process of creating a script and conducting an analytical study of a dataset using the DeepMIMO emulator. An advertorial attack was carried out using the FGSM method to maximize the gradient. A comparison is made of the effectiveness of binary classifiers in the task of detecting distorted data. The dynamics of changes in the quality indicators of the regression model were analyzed in conditions without adversarial attacks, during an adversarial attack and when the distorted data was isolated. It is shown that an adversarial FGSM attack with gradient maximization leads to an increase in the value of the MSE metric by 33% and a decrease in the R2 indicator by 10% on average. The LightGBM binary classifier effectively identifies data with adversarial anomalies with 98% accuracy. Regression machine learning models are susceptible to adversarial attacks, but rapid analysis of network traffic and data transmitted over the network makes it possible to identify malicious activity.

## I. Introduction

The widespread proliferation of the latest generation wireless networks, the development of millimeter wave (mmWave) technologies, massive MIMO (massive Multiple Input Multiple Output) antenna systems and, as a consequence, the increased level of data transmitted over the network from many users entails problems of ensuring network security. Modern machine learning (ML) models are widely used to analyze network traffic and identify malicious network activity, but deep learning models themselves can be vulnerable to adversarial attacks that aim to compromise the effectiveness of such models. Adversarial white box attacks are typical for cases in which the attacker has direct access to machine learning models with the ability to examine the source code and architecture. Adversarial black box attacks are typical for cases in which the attacker has the opportunity to test a ready-made model. Intentionally adding specially trained adversarial perturbations to the source data can lead to compromising the quality of the machine learning model.

The rest of the paper is organized as follows. The second chapter provides an overview of existing literature sources on the topic under study. The third chapter presents the theoretical formulation methods for generating advanced examples. The fourth chapter describes the scheme research of the generated data set. The sixth chapter contains research of adversary attacks on regression models of machine learning. The seven chapter contains a conclusion.

## II. Related Work

Many current studies of adversarial attacks are devoted to the problem of classification based on tabular or graphical data. Should be noted that there are practically no publications on the study of adversarial attacks on regression problems, including in the field of wireless networks, which emphasizes the relevance of this research.

The authors of publication [1] perform an adversarial white-box attack on tabular data, successfully deceiving the neural network and reducing its performance.

The study [2] analyzes the robustness of highly parameterized linear models to adversarial attacks with the goal of maximizing the prediction error.

Article [3] examines the robustness of regression coefficients to adversarial examples prepared to "poison" the initial training data of a ML model.

Publication [4] is devoted to the analysis of the vulnerability of regression models of multivariate time series to adversarial attacks. The authors show that the studied models are vulnerable to attacks, which is critical for security.

The study [5] presents two algorithms for performing adversarial attacks on regression models.

The authors of the article [6] note that prepared adversarial examples generated for a white-box attack can be effectively used to perform an adversarial attack on a regression model unknown to the attacker, that is, to perform a black-box attack.

In [7], a study was carried out to detect adversarial attacks on LSTM and temporal convolutional network prediction models based on one-class support vector machine and local outlier algorithms.

The authors of [8] describe a general approach based on perturbation analysis of learning algorithms to perform adversarial attacks on regression models.

The publication [9] explores ways to mitigate the negative impact of adversarial examples on a robust nonparametric regression model.

The authors of the study [10] note the importance of ensuring security in automotive self-organizing networks and

The research was funded by the Russian Science Foundation (project No. 22-71-10124).

explore various options for performing adversarial attacks on regression models and options for protecting against them.

This article will investigate the impact of adversarial attacks on the quality metrics of machine learning models, as well as how to detect such attacks in various simulated MIMO antenna propagation scenarios.

III. METHODS FOR GENERATING ADVANCED EXAMPLES

An adversarial dodging attack in the case of a classification task is an attack in which the attacker sets the task of incorrectly classifying an object, and it does not matter how exactly the object will be classified and what incorrect class it will be assigned to. In the case of a regression problem, an evasion attack is to sharply increase the error threshold of the regression model, the predicted value should be as large/smaller as possible to the real value. An adversarial poisoning attack is a type of attack performed at the time of training artificial intelligence models associated with mixing "poisoned" data into the training data set. Analyzes modifications of machine learning models by poisoning training data with quantitative risk assessment in the development of artificial intelligence systems. Let's look at basic approaches for generating adversarial samples that can be used to attack machine learning models built on tabular data. The most popular approach is to use the fast gradient sign method.

*A. Fast Gradient Sign Method (FGSM)*

The idea behind this method is that it calculates the gradients of the loss function with respect to the original data, and then uses the sign of the gradients to create a new "poisoned" image that maximizes the machine learning model's J loss:

$$x' = \varepsilon * \text{sign}(\nabla_x J(\theta,x,y)), \quad (1)$$

where $\varepsilon$ is the minimum noise level, $\theta$ is the neural network model, $\text{sign}(\nabla_x J(\theta,x,y))$ is the sign of the gradient, $\nabla_x$ is the gradient, x is the source data, y is the target value for x.

*B. Distance-based attack*

This method is to minimize the distance between the object and the synthetic recording with different output labels. The peculiarity of this approach is the preliminary grouping of adversarial samples in accordance with quasi-identifiers and setting the corresponding secret attribute as the most common value (mode).

*C. Low Profile Algorithm*

This method [11] is to minimize the weighted norm of the disturbance vector on the features of tabular data while maximizing the proportion of examples $x \in X$ with false answers at the output.

IV. GENERATION AND EXPLORATION OF DATA SETS OF MASSIVE MIMO NETWORKS

To generate massive MIMO network datasets from Remcom's precise 3D ray tracing, we used the DeepMIMO framework [12]. The scenario "Boston5G_28" is considered - an open space scenario created based on the center of Boston, with buildings of varying heights. One base station (BS) was recorded on the street at a height of 15 m, equipped with an omnidirectional antenna. The user arrays (UE) are two antenna grids with a total number of 965,090 users, located at a height of 2 m, the distance between users is 37 cm. The standard operating frequency of the emulation is 28 GHz. Each user consists of one omnidirectional antenna. The distance between ray tracing angles is 0.25 degrees. Concrete and wet earth are used as materials for buildings and terrain respectively. The signal propagation model is such that each channel path can go through a maximum of 4 reflections before the base station signal reaches the receiver (user). The bandwidth is set to 0.1 MHz.

The user layout is designed so that most of the users are cut off from the base station in accordance with the topology of the emulated city segment. You can dynamically monitor changes in combined signal losses along the propagation path from the source (base station) to end users, taking into account the architecture of the emulated city segment and signal reflections. For the generated data set, the coordinates of the sender and recipients, the matrix of the sender and recipient channels, as well as various numerical characteristics of signal propagation are available. After performing the scenario calculations, the following features were selected in the final data set:

1. X coordinate – coordinate on the user's X axis relative to the emulated area.

2. Y coordinate – coordinate on the user's Y axis relative to the emulated area.

3. Distance – distance between the base station and each user, in meters.

4. Pathloss – combined losses along the channel path between the sender and the recipient ("attenuation" of the antenna signal), in decibels relative to 1 milliwatt.

5. DoA_phi – azimuthal angle of arrival, in degrees.

6. DoA_theta – zenith angle of arrival, in degrees.

7. DoD_phi – azimuthal departure angle, in degrees.

8. DoD_theta – zenith angle of departure, in degrees.

9. Phase – phase of the signal propagation path, in degrees. 10. Power – signal strength upon reception, in watts.

11. Time of arrival – time of receiving the signal, in seconds.

12. Line of Sight (LoS) – signal status, taking one of three values from {-1, 0, 1}.

(LoS = 1): Line of sight path exists. (LoS = 0): Only non-line-of-sight paths exist, with the line-of-sight path being blocked. (LoS = –1): There are no paths between the transmitter and receiver (total blocking).

V. RESEARCH OF THE GENERATED DATA SET

The resulting dataset contains 105,842 records, with 40,387 users within the base station's line of sight (LoS = 1) and 65,455 users outside the base station's line of sight (LoS = 0). The pathloss metric—combined losses along the channel path—is one of the key metrics for assessing the quality of the latest generation wireless networks and indicates how effective the current network topology is. The pathloss value can be predicted based on available data on the state of the network when transmitting a signal between the base station and a large array of users. An adversarial attack on a regression model should dramatically increase or decrease the predicted value relative to the original value of the target column. To perform an adversarial attack on a signal loss prediction model, it is advantageous for an attacker to dramatically increase the predicted value. Attackers can attack machine learning regression models by poisoning the raw data, causing the combined path loss to skyrocket and users to lose access to the base station under current routing protocols. In the current study, we focus on the problem of generating, detecting, and countering such adversarial attacks.

On the Fig. 1 shows scatterplots and histograms for the combined path loss and the significance of the features for prediction.

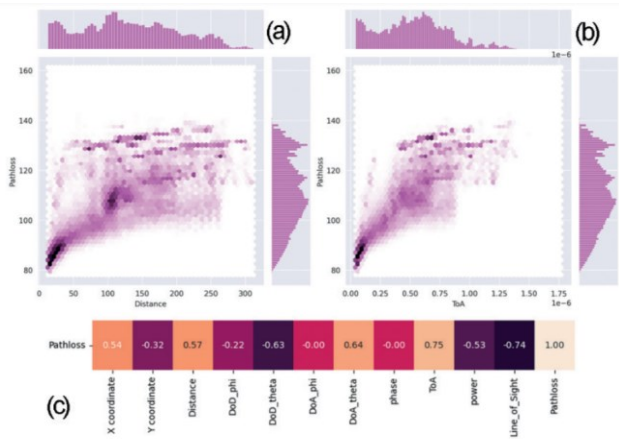

Fig. 1. Histograms of the distribution of the Pathloss feature depending on the user's distance to the base station (a) and depending on the signal arrival time (b), fragment of the correlation matrix (c).

From Figures 1(a) and 1(b), we can visually identify three peaks of high signal loss depending on the distance of the user to the base station and depending on the signal arrival time. Figure 1(c) shows a fragment of the correlation matrix showing a strong direct dependence of the Pathloss feature on the Time of arrival, DoA_theta and Distance features and a strong inverse dependence on the Line of sight, DoA_theta and Power features. Indeed, an increase in signal acquisition time leads to an increase in the combined losses along the channel path. For users within the line of sight of the base station, the combined path loss of the channel is reduced due to the absence of signal reflections along its propagation path.

## VI. RESEARCH OF ADVERSARY ATTACKS ON REGRESSION MODELS OF MACHINE LEARNING

The data set obtained in Section 2 is divided in a ratio of 40:40:20 into a training set for training the regression model, a sample for data poisoning when performing an adversarial attack, and a test set for data validation. Varying all elements of the gradient sign allows you to control the "direction" of the error. On the Fig. 2 shows how the predicted value of the combined pathloss signal changes depending on the sign of the gradient.

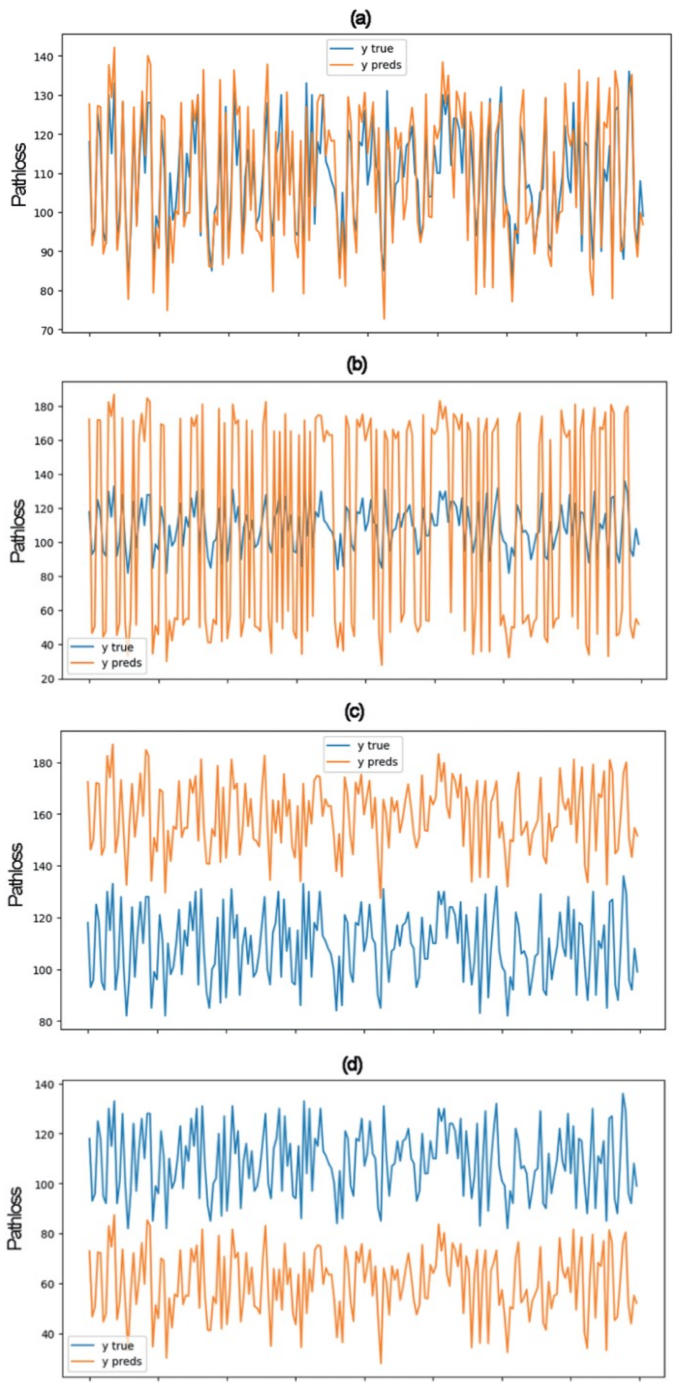

Fig. 2. Fragment of the test data set in various scenarios: (a) – trained linear regression model, (b) – FGSM attack with gradient sign fluctuation, (c) – FGSM attack with gradient sign maximization, (d) – FGSM attack.

The current study examines three main scenarios for investigating adversarial attacks on tabular data:

1. Scenario for training a regression model without third-party interventions (Undefended Model). Let's train the LinearRegressor regressor for the task of predicting combined pathloss losses using the quality assessment metric Mean Squared Error (MSE) and R2. The linear regression model showed good accuracy when solving the problem of predicting the pathloss indicator based on other features. When building a neural network architecture, gradient descent converges at a local extremum point, so the general algorithm for training a regression model is as follows:

1.1 Linear regression was trained from the sklearn library based on the least squares method.

1.2 The resulting weights and free coefficient (shift) were used to initialize a neural network with one linear layer and without an activation function using the pytorch library.

1.3 The constructed neural network has been tested.

1.4 The quality metrics of the regression model were calculated.

2. Scenario for poisoning source data for training based on generative adversarial networks (Attacked Model). Let's perform an adversarial FGSM attack with varying the neighborhood index $\varepsilon = [1^{-10}, 1^{-9}, 1^{-8}, 1^{-7}]$ and the fraction of attacked data fract = [0.2, 0.4, 0.6, 0.8, 0.95, 0.99999].

Figure 3 shows the dependence of quality metrics on the size $\varepsilon$ of the neighborhood for the trained linear regression model. A value of $\varepsilon = 1-7$ and higher leads to a sharp increase in the values of the MSE metric and a decrease in the values of the R2 metric, which is inappropriate when conducting an adversarial attack, because a very strong model deviation will be considered an outlier or an anomaly in the data.

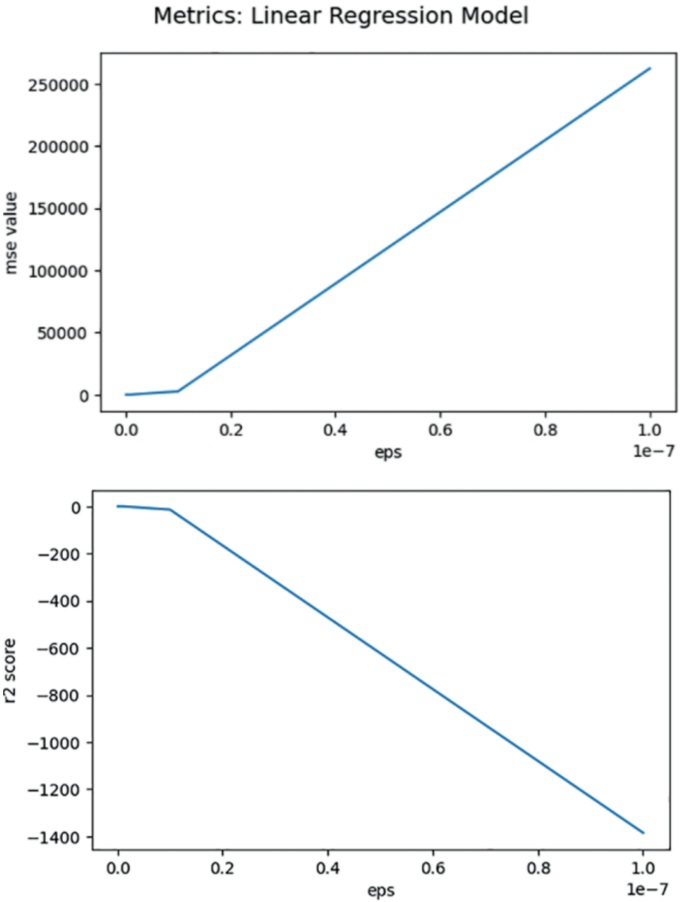

Fig. 3. Dependence of the values of the MSE and $R^2$ score metrics on the size $\varepsilon$ of the neighborhood.

As a result of studies on varying FGSM parameters, we can conclude that the linear regression model is most vulnerable to the FGSM attack with maximizing the sign of the gradient with parameters $\varepsilon = 1^{-10}$ and fract = 0.99999; in other configurations, deviations in metrics are insignificant.

3. Scenario for detecting and countering adversarial attacks on source data (Secured Model). The poisoned data sets obtained in the second scenario were used to train the LightGBM, CatBoost and XGBoost classifiers to solve the binary classification problem: normal data (benign data) labeled "0" or poisoned data (malicious data) labeled "1". The optimal parameters of the classifiers were selected using the GridSearchCV tool.

Table 1 presents the results of comparing the three classifiers. To train the classifiers, a data set is randomly selected in which the poisoned and normal data are proportionally balanced.

TABLE I. COMPARISON OF BINARY CLASSIFIERS FOR ANOMALY DETECTION

| Classifier | $\varepsilon = 1^{-10}$, fract = 0.6 | | |
| --- | --- | --- | --- |
| | *Precision* | *Recall* | *F1-score* |
| LGBMClassifier (max_depth=20, n_estimators=500, num_leaves=20, subsample=0.7) | 0.9835 | 0.9833 | 0.9834 |
| CatBoost (depth'=4, 'learning_rate'=0.02, 'iterations'=100) | 0.9777 | 0.9646 | 0.9703 |
| XGBoost (n_estimators=500) | 0.9828 | 0.9816 | 0.9822 |

The best results in detecting adversarial anomalies are shown by the LightGBM classifier with the parameters max_depth=20, n_estimators=500 num_leaves=20, and subsample=0.7. Based on the results of the classifier's work on the test data, we will remove detected adversarial examples from the data set and obtain a reduced data set of 5029 records, on which we will re-evaluate the quality of the regression model.

At each of the three stages, the main metrics of the quality of regression models were calculated. Table 2 shows the dynamics of changes in the quality metrics of the linear regression model depending on the scenario under study. Performing an adversarial FGSM attack with maximizing the sign of the gradient and parameters of the neighborhood indicator $\varepsilon = 1^{-10}$ and the fraction of attacked data fract = 0.99999 increases the value of the MSE metric by an average of 33% and reduces the value of the $R^2$ metric by an average of 10%. The LightGBM binary classifier with selected optimal hyperparameters successfully detects records with adversarial anomalies in tabular data with an accuracy of 98%, the isolation of which allows us to restore the regression model metrics to their original values.

TABLE II. THE DYNAMICS OF CHANGES IN QUALITY METRICS LINEAR REGRESSION MODEL

| Scenario | $\varepsilon = 1^{-10}$, fract = 0.99999 | |
| --- | --- | --- |
| | *MSE* | *$R^2$* |
| Undefended Model {Linear Regression} | 38.51 | 0.8 |
| Attacked Model {FGSM} | 51.40 ↑ | 0.72 ↓ |
| Secured Model {LightGBM} | 37.55 ↓ | 0.80 ↑ |

VII. CONCLUSION

As part of the study, tabular data for a wireless network segment scenario was generated based on the DeepMIMO emulator; the construction of adversarial examples was completed in order to maximize the predicted value of the combined signal losses from the base station to end users; a binary classifier was trained to recognize poisoned data; shows the dynamics of changes in quality metrics of a linear regression model in applications of 6G wireless networks. While machine learning regression models are vulnerable to adversarial attacks, timely intelligent analysis of network traffic and data transmitted over the network can detect malicious network activity in the latest generation wireless network segment.